# Chromatic annuli formation and sample oxidation on copper thin films by femtosecond laser

*Shutong He,†,‡, Salvatore Amoruso,‡ Dongqing Pang,† , Chingyue Wang† and Minglie Hu,\* †*

†Ultrafast Laser Laboratory, Key Laboratory of Opto-electronic Information Technical Science of Ministry of Education, College of Precision Instruments and Opto-electronics Engineering, Tianjin University, Tianjin 300072, China

**ABSTRACT:** We report an experimental investigation on the irradiation of copper thin films with high repetition rate femtosecond laser pulses (1040 nm, 50 MHz), in ambient air and liquid water. We observe a novel, striking phenomenon of chromatic copper oxides (CuO and $Cu_2O$) annuli generation. The characteristic features of the chromatic copper oxide annuli are studied by exploiting micro-Raman Spectroscopy, Optical and Scanning Electron Microscopies. In the case of irradiation in water, the seldom investigated effects of the immersion time, $t_w$, after irradiation with a fixed number of pulses, is analyzed, and an intriguing dependence of the color of the chromatic annuli on $t_w$ is observed. This remarkable behavior is explained by proposing an interpretation scenario addressing the various processes involved in the process. Our experimental findings show that $Cu_2O$ nanoparticles (size of ≈20 nm) and $Cu_2O$ nanocubes (nanocube edges of

≈30, ≈60 nm) with extremely fine sizes can be effectively generated by exploiting high repetition rate laser-assisted oxidation.

## 1. Introduction

Metal oxide nanoparticles (NPs) are attracting considerable attention for their potential applications in different fields, e.g. opto- and nano-electronics, gas sensing, catalysis, and so forth. Among various metal oxide, copper oxide NPs are particularly interesting thanks to a range of useful physical properties. Copper has two stable oxides, CuO and $Cu_2O$. CuO is narrow bandgap (≈1.2 eV), *p*-type semiconductor with useful photocatalytic and photovoltaic properties, and CuO NPs are increasingly used in catalysis, gas sensors, heat transfer fluids, solar energy, surface enhanced Raman spectroscopy and as an antimicrobial agent.[1,2] $Cu_2O$ is a *p*-type semiconductor with particular optical and magnetic properties (direct band gap of 2.17 eV), whose NPs are of interest for several applications.[3,4] For example, $Cu_2O$ is a potential photovoltaic material and is appropriate for solar energy conversion,[5–8] and a stable catalyst for photoactivated splitting of water or other liquids under visible light irradiation;[9–11] Moreover, large area porous $Cu_2O$ has been used for super hydrophobic yellow-red phosphors.[12] Finally, both $Cu_2O$ and CuO nanospheres can be used for gas sensing.[13]

Many chemical methods have been devised to produce $Cu_2O$ film or $Cu_2O$ nanostructures (e.g. nanocubes), such as electrodeposition, thermal relaxation, complex precursor surfactant-assisted route, vacuum evaporation, etc.[14–20] However, the production of fine $Cu_2O$ nanocubes (nanocube edge < 100 nm) is difficult, since appropriate choice of chemical reagents, such as capping agents, is extremely important and complicated procedures are involved in the process. In addition, the production needs a relative long time, e.g. more than 2 hours are required for seed-mediated synthesis.[21–25]

Laser-assisted oxidation is an alternative for classical chemical methods. For example, laser-assisted oxidation with Nd:YAG cw-laser was exploited for the coloration of titanium plates.[26,27] Fine $Cu_2O$ NPs ($\approx$10 nm) were produced by pulsed laser ablation in liquid induced by pulses with nanosecond duration.[28] The availability of high repetition rate femtosecond (fs) laser sources is opening up novel approaches in laser material processing and can possibly offer advantages also in laser-assisted oxidation. For example, exposure of metal thin films to high repetition rate ultrafast laser pulses have evidenced particular oxidation reactions facilitating the growth of *m*-$MoO_2$ and *o*-$Mo_4O_{11}$.[29]

In this work, we investigate laser-assisted oxidation of copper thin films induced by high repetition rate fs pulses (1040 nm, 74 fs pulse duration, 50 MHz) provided by an Yb-doped large mode area photonic crystal fiber (LMAPCF) laser source. After exposure, the irradiated samples are characterized by micro-Raman Spectroscopy ($\mu$RS), Optical Microscopy (OM) and Scanning Electron Microscopy (SEM). Two sets of experiments were carried out. First, Cu films with different thicknesses (50, 100 and 200 nm) were irradiated in air, addressing the influence of film thickness on the laser-assisted oxidation process. We observe the formation of chromatic rings around the ablated craters composed of both CuO and $Cu_2O$ NPs, with CuO closer to the beam spot location and $Cu_2O$ at larger distance. The radius of the copper oxide rings varies with the film thickness, which shows a striking influence of the formation of the two oxides that is related to the thermal history of the sample. In particular, the Cu films with reduced thicknesses show better diversity of copper oxides and more sensitivity to the oxidants during laser irradiation. Consequently, we choose 50 nm thickness Cu films for the second set of experiments which were carried out by exposing the sample in water environment. With the aid of water, we have observed a better efficiency of the generation of copper oxides. The surface oxidation process was analyzed as a function of the time when the sample remained immersed in water (immersion time, $t_w$) which is seldom considered before. We observed an interesting effect of the immersion time in water, and an efficient formation of $Cu_2O$ nanocubes with ultrafine size (nanocube edge ≈30 nm) was addressed in appropriate conditions.

## 2. Experimental methods

Figure 1 reports a schematic of the experimental setup. An Yb-doped LMAPCF femtosecond (fs) laser amplifier delivered linearly polarized pulses with a duration of ≈74 fs at a central wavelength $\lambda$=1040 nm and a repetition rate of 50 MHz. A combination of half waveplate ($\lambda/2$) and polarization beam splitter (PBS) was used to adjust the output pulse energy. The beam was focused by an objective lens (Mitutoyo M Plan Apo NIR 20×, NA=0.4) at a working distance of 20 mm. The $1/e^2$ laser spot size diameter on the target surface was $w_0$≈10 $\mu$m. The target surface was illuminated by a light emitting diode (LED) and observed in real time by exploiting a charge coupled device (CCD) camera.

The Cu thin films were deposited on BK7 substrates. A first set of experiments were carried out in air. Samples with three different thicknesses (*H*) of the Cu thin film, namely ≈50 nm, ≈100 nm and ≈200 nm were used to investigate the influence of *H* on laser induced oxidation. The average fluence at the focus was ≈0.13 J/cm². Preliminary experiments showed the progressive formation of chromatic rings extending over a region larger than the area irradiated by laser (see Figure 1(b)) and whose features change as a function of the laser irradiation time. Therefore, in this investigation we fixed an irradiation time of 1 min (3 billion pulses), which corresponds to experimental conditions for which the chromatic rings clearly appears on the sample surface.

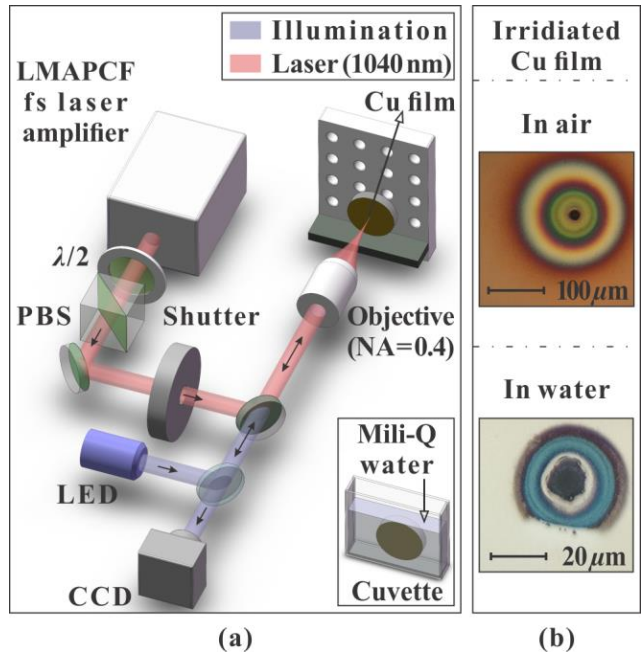

**Figure 1.** (a) Schematic of the experimental setup. The lower inset shows the target configuration used for experiment in water. (b) Optical microscope images of the irradiated Cu films, in air (upper panel) and water (lower panel) evidencing the formation of chromatic rings.

As for the experiments in water, the Cu thin film sample was put in a cuvette with a large pure fused quartz window holding Mili-Q water (see Figure 1(a), inset). Cu thin films with a thickness $H \approx 50$ nm were used and laser irradiated for a time of 200 ms (10 million pulses) at an average fluence at the focus of $\approx 0.13$ J/cm$^2$ by considering the absorption of water, i.e. at the same fluence as that in air. In such conditions, the effects of the overall time interval the Cu film sample was immersed in water ($t_w$) was analyzed. The effect of $t_w$ was seldom investigated in previous studies of laser induced oxidation of metals. Interestingly, our experimental findings suggest that the features of the chromatic rings in presence of water strongly depend on the immersion time $t_w$. In particular, we carried out experiments with $t_w$ varying from 0 min to 5 min, and investigated the color modification of the chromatic annuli.

The optical properties of the laser-exposed areas (e.g. the colors of the chromatic rings) were examined by an Optical Microscope (OM, Hirox KH-1000 microscope), while a Scanning Electron Microscope (SEM, Nova NanoSEM 430) was exploited to analyze the associated morphological modifications of the samples. The materials formed in the various chromatic areas formed around the focal region were studied by micro-Raman Spectroscopy ($\mu$RS, Renishaw inVia).

## 3. Results and discussion

Prior to illustration of experimental results, it is worth considering some important aspects peculiar to our experimental conditions of high repetition rate fs laser pulses irradiating a copper thin film on a glass substrate. At the wavelength of the laser pulses ($\lambda$=1040 nm), the copper optical absorption depth is $\approx$12 nm.[30] Hence, even for the smaller film thickness of 50 nm more than 99% of the pulse energy is absorbed within the film, and the fraction transmitted to the glass substrate is negligible. Moreover, the heat accumulated in the target during irradiation is transported more effectively along the film than through the BK7 glass substrate, due to the diverse thermal conductivity $k$ ($k_{Cu}\approx$350 Wm$^{-1}$K$^{-1}$; $k_{BK7}\approx$1.1 Wm$^{-1}$K$^{-1}$) and diffusivity $D$ ($D_{Cu}\approx$1.1×10$^{-4}$ m$^2$s$^{-1}$; $D_{Cu}\approx$5.3×10$^{-7}$ m$^2$s$^{-1}$) of the two materials.[30] Therefore, thanks to the high thermal conductivity of copper, heat transfer in the areas surrounding the laser beam spot location is expected for long irradiation time, as indeed observed in Figure 1(b).

Study on the irradiation of thin films on a transparent substrate with fs laser pulses indicate that the damage threshold reduces with the film thickness.[31,32] For copper films with thickness $t$ irradiated by 50 fs laser pulses at 800 nm, Wang and Gallais calculated single shot laser induced damage threshold of ≈0.5 J/cm$^2$ for $t$=50 nm, ≈1.1 J/cm$^2$ for $t$=100 nm and a plateau value of ≈1.5 J/cm$^2$ for $t$≥200 nm. Similar values are expected at 1030 nm due to the very similar reflectivity of copper at the two wavelengths.[32] Therefore, we can assume that the laser fluence of ≈0.13 J/cm$^2$ used in the present study is well below the single shot melting and ablation thresholds of the copper films, thus any effect of pulse shielding due to material ablation during the laser irradiation can be considered negligible in our experimental conditions.

In the following, we will first discuss the results of the experiments carried out in air, in section 3.1. Then, in section 3.2, we will illustrate how the presence of a water environment affects the laser processing of the copper film.

**3.1. Laser irradiation of copper films in ambient air.** Figure 2 reports OM micrographs of the Cu thin films with thickness $H$≈50 nm (a), $H$≈100 nm (b) and $H$≈200 nm (c), respectively, after irradiation with 3 billion fs laser pulses at an average fluence of ≈0.13 J/cm$^2$. The OM images clearly evidence the formation of chromatic annular regions on the Cu thin films. In Figure 2, the coordinate axes of the images represent the distance from the center of the laser beam spot, which allows assessing the radius of the different chromatic annuli, $r$. The colors of the 50 nm thick Cu film surface allows to easily distinguish various regions with color changing from taupe brown, to grayish green and to sky blue as $r$ increases, while the copper film surface can be again recognized at $r$>110 $\mu$m (see Figure 2(a)). Different chromatic rings are also identified for the film with $H$≈100 nm, in Figure 2(b). Moreover, the Cu film with $H$≈200 nm presents a larger number of colored regions with alternating pink and green chromatic rings (see Figure 2(c)). This, in turn, suggests an apparent rise of the number of the chromatic annuli as the Cu film thickness increases.

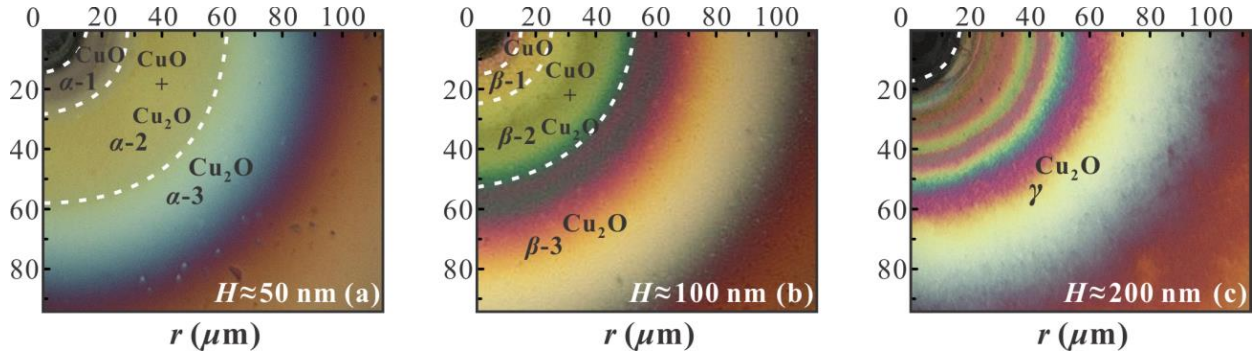

**Figure 2.** OM images of the surface of the Cu thin films with thickness of: (a) $H\approx$50 nm, (b) $H\approx$100 nm and (c) $H\approx$200 nm, after irradiation with 3 billion fs laser pulses at an average fluence of $\approx$0.13 J/cm$^2$ in air. The images illustrate the formation of diverse chromatic annuli on the film surface. The white circles identify different annular regions associated to the predominance of a particular copper oxide and are indicated as $\alpha$-1, $\alpha$-2 and $\alpha$-3 in panel (a), $\beta$-1, $\beta$-2 and $\beta$-3 in panel (b) and $\gamma$ in panel (c), as indicated by $\mu$RS measurements.

The composition of the various regions formed on the Cu film surface was analyzed by $\mu$RS measurements. Different micro-areas of the various annular regions were measured, and Figure 3 reports the typical Raman spectra registered. The observed Raman signals are suggestive of copper oxide formation. In particular, the Raman peaks at 147 and 218 cm$^{-1}$ are assigned to Cu$_2$O, whereas the peaks at 280 and 330 cm$^{-1}$ are characteristic of CuO, while the peak at 617 cm$^{-1}$ is reported both for CuO and Cu$_2$O.[18,14,33–37] This allows one to assess the composition of the observed annular regions formed on the Cu thin films according to the corresponding characteristic Raman spectrum. In particular, we can distinguish the following regions, as indicated on the OM images of Figure 2: $α$-1, $α$-2 and $α$-3 for $H≈50$ nm; $β$-1, $β$-2 and $β$-3 for $H≈100$ nm, and a single region $γ$ for $H≈200$ nm. According to Raman measurements (see Figure 3), regions $α$-1 and $β$-1 are prevalently composed of CuO due to a dominance of the characteristic CuO peak at 280 cm$^{-1}$. Instead, $Cu_2O$ prevails in the regions $α$-3, $β$-3 as indicated by the prominence of the characteristic Raman peaks at 147 and 218 cm$^{−1}$. The intermediate regions $α$-2 and $β$-2 of the Cu films with thickness of 50 and 100 nm shows all the copper oxides Raman peaks in different proportion, suggesting a mixture of CuO and $Cu_2O$. Strikingly, the characteristic Raman spectra registered for the 200 nm thick Cu sample show high intensity peaks at 147 and 218 cm$^{−1}$ and a very weak CuO contribution at 280 cm$^{-1}$, indicating that $Cu_2O$ is mainly formed in the region $γ$ in this case. Interestingly, there is a close correspondence between the three regions with different composition and the chromatic annuli identified in the OM image of Figure 2(a) for the Cu film with $H≈50$ nm. Instead, for $H≈100$ nm the $Cu_2O$ outer annulus $β$-3 presents several colored ring-shaped sub-regions that become even more for $H≈200$ nm. Moreover, the spatial width of these ring-shaped sub-regions tends to increase for larger radius $r$.

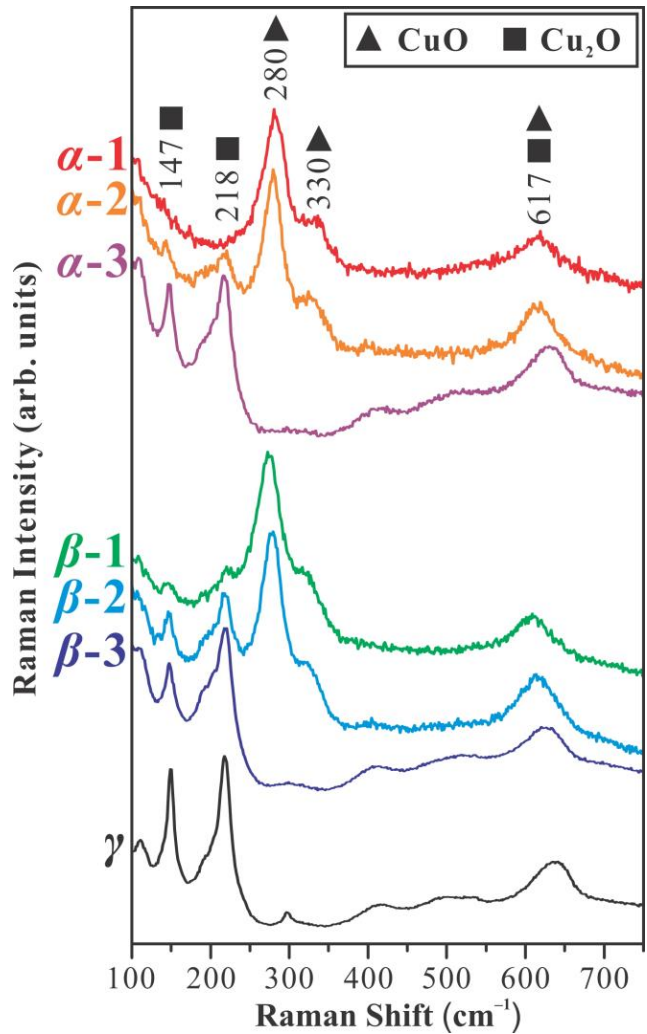

**Figure 3.** Typical μRaman spectra registered in the different annular regions indicated in Figure 2. The spectra are vertically shifted for easiness of comparison.

Previous studies on laser-assisted oxidation of Cu thin films indicated that the induced rise of the temperature $T$ leads to $Cu_2O$ formation on the film surface for $T<300$ °C, while CuO develops for $T>300$ °C.[38,39] A such scenario is pretty consistent with our experimental results. In fact, for $H\approx50$ nm and $H\approx100$ nm, CuO forms in the regions α-1 and β-1 located closest to laser irradiated area where larger temperature is expected. Then, moving to larger distance $r$, a progressive reduction of the temperature eventually leads to the Cu$_2$O regions *α*-3 and *β*-3 after passing through an intermediate area (i.e. *α*-2 and *β*-2) where both phases coexist. The observed spatial variation of the copper oxide composition seems to reflect a spatial distribution of the mean temperature generated by heat conduction going from the irradiated zone towards the periphery. As the film thickness increases, a lowering of the average temperatures is expected due to the increased quantity of material to heat up. Moreover, it is worth observing that, for very thin metallic films, heat conductivity tends to vary with thickness and generally reduces for thinner films.[40] This is in good agreement with our experimental observations of a slightly larger heat affected zone for $H \approx 100$ nm with respect to $H \approx 50$ nm. As for the Cu film with $H \approx 200$ nm, the rather uniform composition dominated by the Cu$_2$O phase suggests that the mean temperature never reaches values higher enough to effectively promote the formation of CuO, which is coherent with its larger thickness. However, it is worth noticing that a "purely thermal" interpretation cannot fully rationalize our experimental findings. In fact, the abundant colors and alternating chromatic rings present in the regions *β*-3 and *γ*, i.e. in a region with rather uniform composition, points to a strict relationship with the peculiar thermal history experienced by the copper thin films under the periodic heating process associated to irradiation with high repetition rate fs pulses. A complete description of such phenomenon is not yet clear at this stage and will deserve further experimental investigation and theoretical analysis. However, even at this stage, it is possible to provide some explanations of the observed spatial variation of composition with film thickness, as described above.

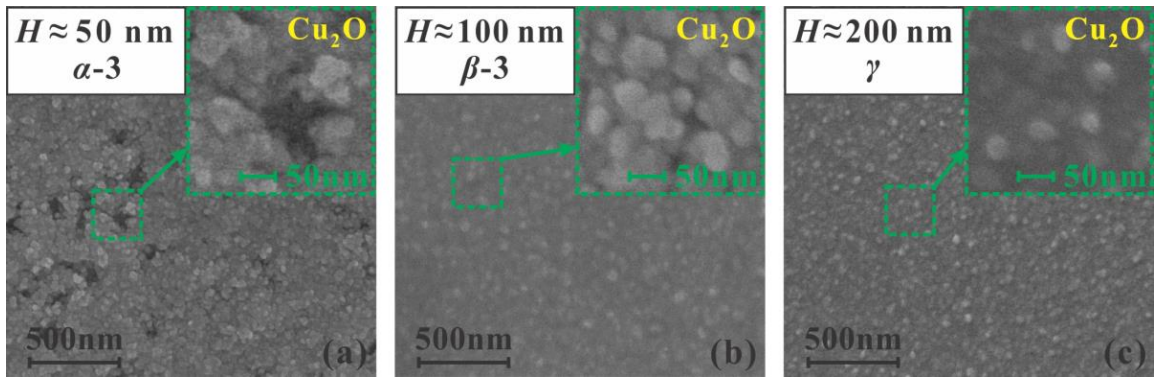

**Figure 4.** SEM micrographs of the regions occupied only by Cu$_2$O, i.e. (a) $\alpha$-3, (b) $\beta$-3 and (c) $\gamma$. The upper-right insets correspond to the areas marked by dashed squares in green.

Figure 4 reports SEM micrographs of the three Cu thin films surface registered in the regions where only Cu$_2$O is generated, i.e. $\alpha$-3, $\beta$-3 and $\gamma$. The SEM images of Figure 4 evidences a nanostructured morphology of the processed surface, and a progressive reduction of the average size of the Cu$_2$O nanoparticles (NPs) as the film thickness increases. Moreover, the SEM micrograph of the thinner film ($H\approx$50 nm) shows the presence of a number of holes, one of which is marked by the dashed green box in Figure 4(a). The formation of the holes can be due to the very small thickness of the copper film, since this effect is observed for the thicker ones. However, SEM analysis does not evidence any feature related to the different chromatic rings. In fact, the average NPs size and the surface morphological features remain rather uniform over the several colored annular sections of the $\gamma$ region.

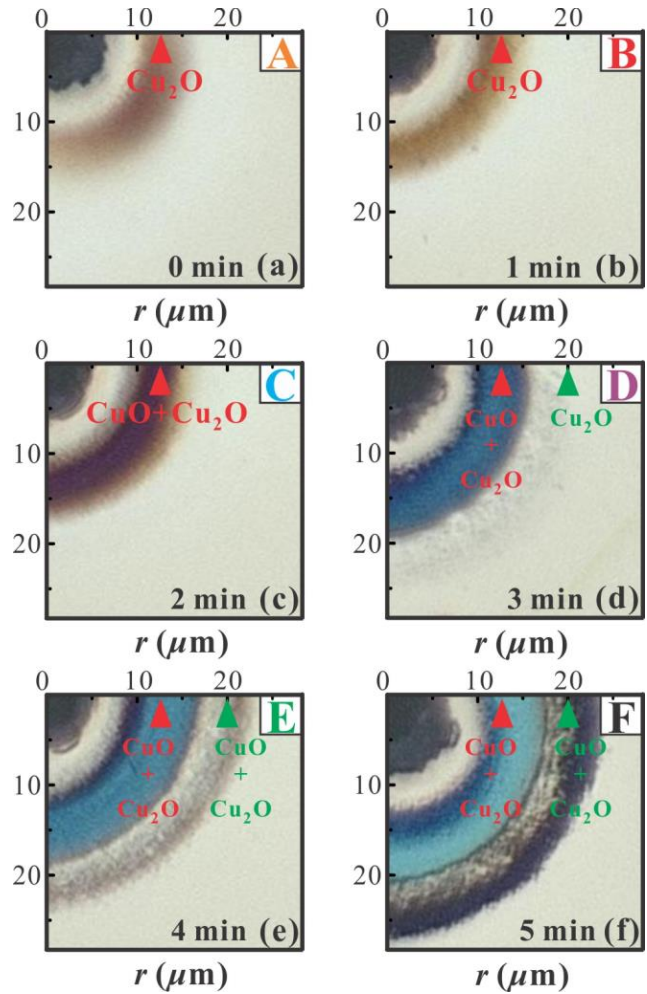

**Figure 5.** OM micrographs of the chromatic annuli formed on Cu thin films in water for different values of the immersion time $t_w$ varying from 0 to 5 min for (a)–(f), respectively. The upper-right capital letters from A to F represent the name of the samples displayed in the panels (a)–(f), respectively. The red arrows at $r=12.5$ μm in panels (a)–(f) indicate the chromatic areas, and the green arrows at $r=20$ μm in panels (d)–(f) point out the outer black annuli.

**3.2. Laser irradiation of copper films in water.** We turn now to the discussion of the experiments on laser processing of the copper thin film in liquid water environment. In particular, 50 nm thick Cu films were laser irradiated in water for 200 ms (10 million pulses) at an average laser fluence of ≈0.13 J/cm$^2$ which is the same with the fluence in air, and then left for a time interval $t_w$ in the water, after the irradiation. Figure 5, reports the OM micrographs of the sample for different values of the immersion time $t_w$, from 0 up to 5 min. The value $t_w$=0 corresponds to the case of a sample kept in the water only during laser irradiation. In the OM images of Figure 5, the coordinate axes represent the distance from the center of the laser beam spot, $r$. The samples are identified with the letters A-F according to the corresponding $t_w$ value, as indicated in the upper-right corner of each OM image in Figure 5.

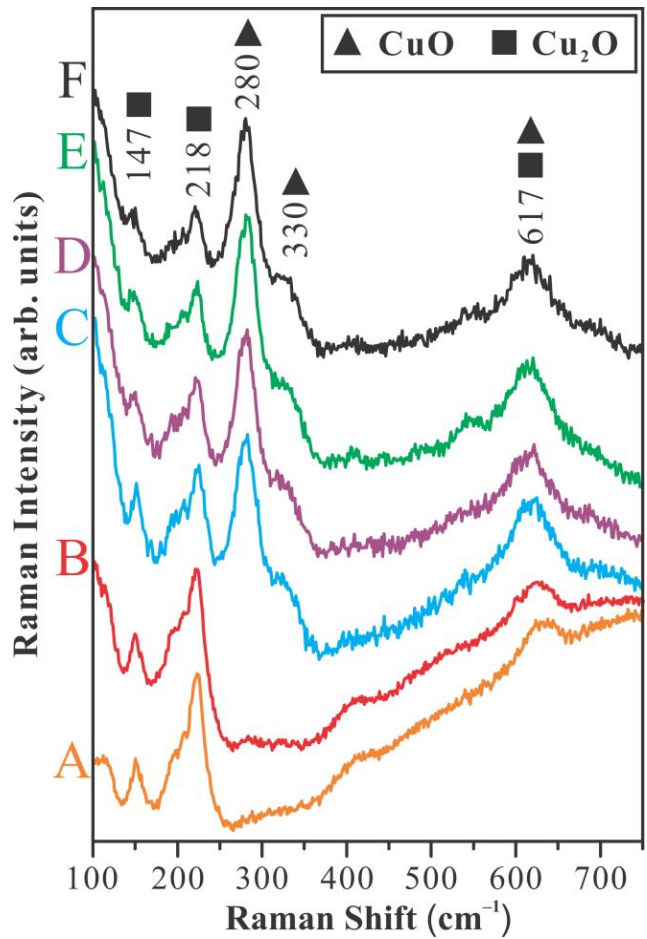

**Figure 6.** μRaman spectra corresponding to the micro-areas at $r≈12.5$ μm indicated by the red arrows in Figure 5 for the samples from A to F. The spectra are vertically shifted to facilitate the comparison.

At low values of the immersion time ($t_w$<3 min), we observe the distinct formation of a chromatic annular region with an average radius $r≈12.5$ μm, which is pointed by red arrows in the OM images of Figure 5. The color of this ringed area varies distinctly as a function of $t_w$; passing from sample A to F it is: brown, yellowish brown, violet, navy, azure and cyan, respectively. At $t_w$=3 min (sample D) an outer black annulus at $r≈20$ μm appears, which then becomes progressively darker at larger values of $t_w$, for samples E and F. The characteristic radius of this additional feature is marked by green arrows in the panels (d)-(e) of Figure 5.

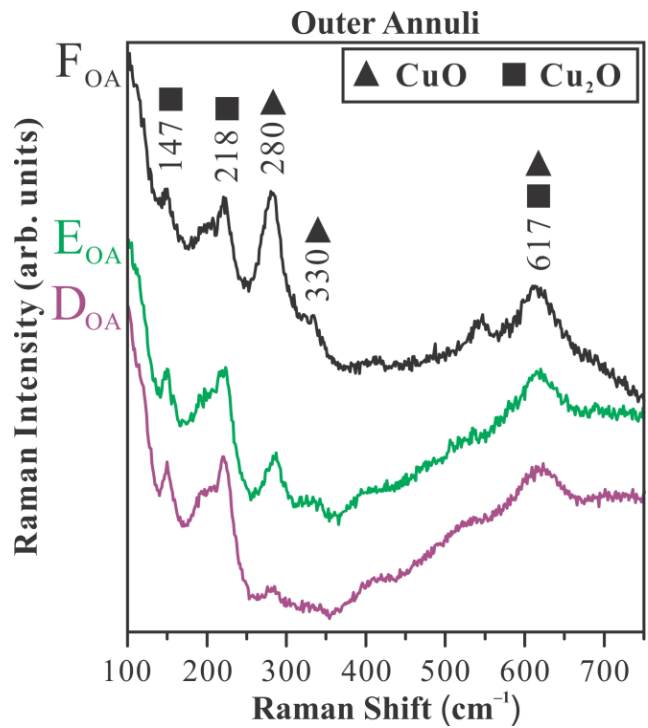

**Figure 7.** $\mu$Raman spectra corresponding to the outer annuli at $r\approx20$ $\mu$m indicated by the green arrows in Figure 5, corresponding to regions $D_{OA}$, $E_{OA}$ and $F_{OA}$, respectively. The spectra are vertically shifted to facilitate the comparison.

The composition of the two annular regions was analyzed by $\mu$RS measurements. For easiness of identification, in the following we will refer to the inner annulus at $r\approx20$ $\mu$m as A-F, and the outer one at $r=20$ $\mu$m on samples D–F will be indicated as $D_{OA}$–$F_{OA}$. Figure 6 reports characteristics $\mu$Raman spectra corresponding to the micro-areas at $r\approx12.5$ $\mu$m indicated by the red arrows in Figure 5 for the samples A-F. The Raman spectra of samples A and B present negligible signature of CuO peaks, thus suggesting that the brownish ring of Figures 5(a) and 4(b) formed at $t_w\leq 2$ min is mainly in the $Cu_2O$ phase. As $t_w$ increases to values larger than 2 min, CuO peaks appear and progressively intensify with respect to the $Cu_2O$ Raman signals. This, in turn, suggests a progressive rise of the CuO fraction present in the chromatic ring at $r\approx12.5$ $\mu$m as $t_w$ increases, which is related to the chromatic variation of the inner annulus. As CuO starts forming for $t_w=3$ min, the annulus color turns towards blue, and the resulting violet color in Figure 5(c) results from a mixture of brown (related to $Cu_2O$) and blue (related to CuO). Then, the gradual enhancement of the CuO fraction as $t_w$ increases, as indicated by the $\mu$Raman spectra of Figure 6, leads to a progressively more bluish coloration of the annulus. Then, for $t_w=5$ min (sample F), the chromatic ring is mainly formed by CuO.

Figure 7 reports the characteristic $\mu$Raman spectra registered in micro-areas of the outer rings at $r\approx20$ $\mu$m, i.e. in samples $D_{OA}$–$F_{OA}$. The evolution of the Raman features for the outer rings show a similar dependence with $t_w$. At lower irradiation time (i.e. for $t_w=3$ min), the Raman spectrum of the ring $D_{OA}$ is suggestive of a region merely composed by $Cu_2O$. As $t_w$ increases, CuO gradually generates in the annulus $E_{OA}$, and eventually a consistent fraction of CuO is formed in the annulus $F_{OA}$.

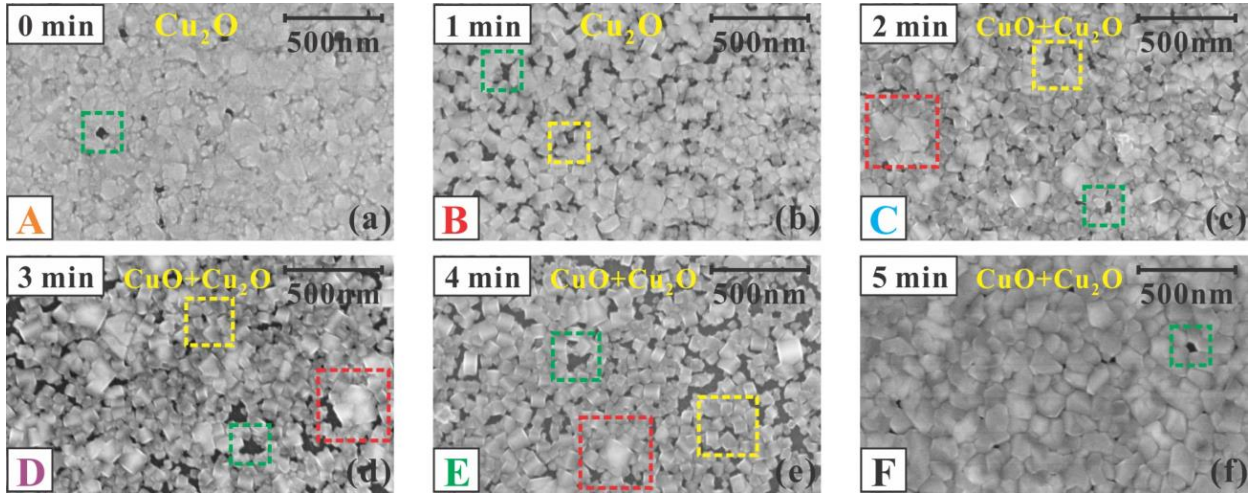

**Figure 8.** SEM micrographs of micro-areas at $r \approx 12.5$ $\mu$m indicated by the red arrows in Figure 5 for samples from A to F. The green dashed squares identify holes where the underlying substrate surface can be recognized, the yellow dashed squares mark $Cu_2O$ nanocubes, and the red dashed squares indicate the places where CuO forms. The yellow labels in the upper center of the images indicates the composition of the sample as obtained by the $\mu$Raman spectra of Figure 6.

Figure 8 reports SEM micrographs registered in the chromatic annuli at $r \approx 12.5$ $\mu$m for samples A-F. In this regions, the surface of the Cu film is initially oxidized to $Cu_2O$ at the end of laser irradiation, as indicated by the $\mu$Raman analysis (see Figure 6, sample A). The SEM image of Figure 8(a), for sample A ($t_w$=0) shows the appearance of $Cu_2O$ nanostructure mainly composed by still tiny and irregular NPs. Then, the SEM image of sample B ($t_w$=1 min) in Figure 8(b) evidences that $Cu_2O$ nanocubes starts forming after immersion in water for 1 min, as illustrated by the dashed yellow square. The two stable copper oxides have different lattice structures: $Cu_2O$ is cubic and CuO is monoclinic. Hence, the formed nanocubes should be $Cu_2O$,[28] in agreement with the μRaman spectra of Figure 6. At immersion time $t_w$ larger than 2 min, CuO is progressively formed (see μRaman spectra in Figure 6). This leads to the formation of relatively bigger particles, as those observed in the dashed red square area in Figure 8(c), whose irregular shape is caused by the monoclinic CuO lattice structure. According to μRaman spectra of Figure 6, these particles are likely composed by CuO or $Cu_2O$/CuO mixture. Comparing SEM images of Figs. 8(d) and 8(e), one can observe that the amount of the $Cu_2O$ nanocubes (see yellow dashed boxes) in the sample E is larger than in sample D. Moreover, similar irregular bigger particles are recognized in both cases, as depicted in the dashed red squares. Then, in Figure 8(f) most of the area on the chromatic ring of sample F ($t_w$=5 min) consists of CuO or $Cu_2O$/CuO irregularly-shaped particles.

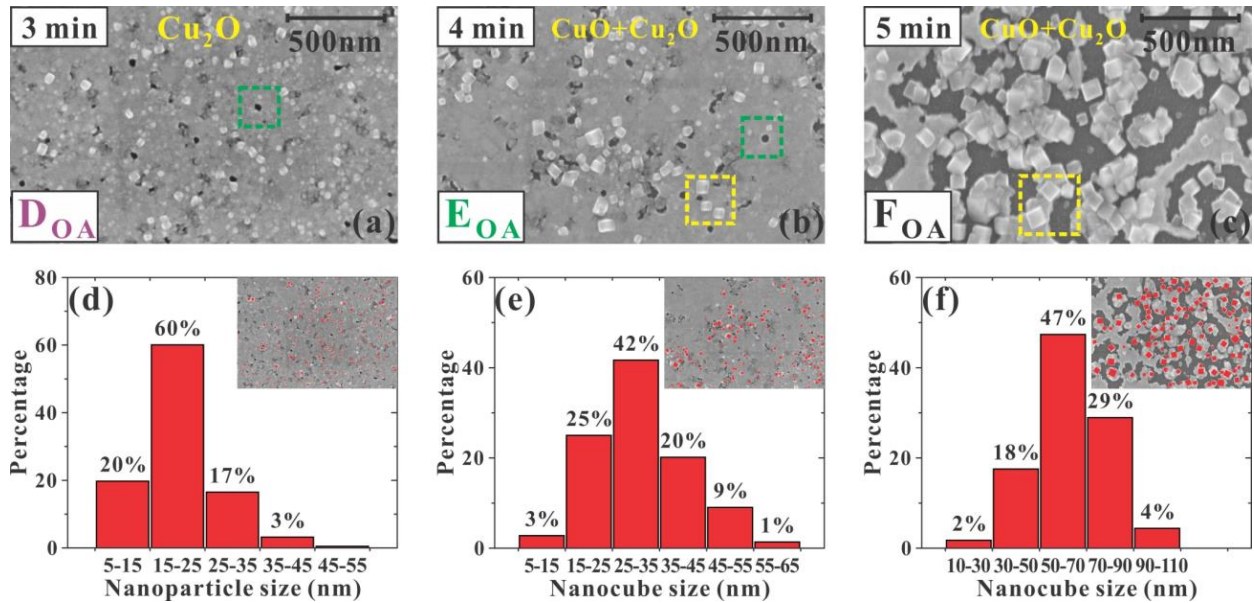

**Figure 9.** Panels (a)–(c): SEM micrographs of the regions $D_{OA}$, $E_{OA}$ and $F_{OA}$, respectively. Dashed squares in yellow in (b) and (c) mark the $Cu_2O$ nanocubes, and dashed squares in green indicate the holes where uncovered substrate surface is present. The yellow labels in the upper center of the SEM images indicate the composition of the sample as obtained by the $\mu$Raman spectra of Figure 7. Panels (d)–(f): Histograms of the size distributions corresponding to (a)–(c), respectively. The upper-right insets shows the NPs/nanocubes selected for the size distribution measurement.

The dashed green squares in the SEM images of Figure 8 identify places where the underlying substrates surface is revealed, i.e. areas without coverage of copper or copper oxide material. Going from sample A to F, is seems that $Cu_2O$ nanocubes form close to the uncovered substrate regions, and we propose that the Cu film provides the original material to generate the copper oxides in these regions. With increasing of immersion time, the size of the uncovered holes grow larger till $t_w$=4 min, and then severely reduces as the fraction of the big CuO and $Cu_2O$/CuO particles increases and the amount of $Cu_2O$ nanocubes drops down.

Figure 9 reports the SEM images of the outer annuli at $r\approx20$ $\mu$m of samples $D_{OA}$, $E_{OA}$ and $F_{OA}$ (upper panels) indicated by the green arrows in Figure 5(d)–5(f). The images show a quite distinct morphology with respect to the chromatic rings at $r$=12.5 $\mu$m. The dashed green squares in Figure 9 identify uncovered holes where the substrates surface is revealed in this case. One can observe that the margins of the holes are smooth and clear and, distinctly from the uncovered places present in the inner rings of Figure 8, are not decorated by NPs. At $t_w$=3 min (sample $D_{OA}$), the presence of several $Cu_2O$ NPs scattered on the Cu film surface is recognized in Figure 9(a). Then, at $t_w$=4 min when CuO starts forming, according to $\mu$Raman spectra of Figure 7, the $Cu_2O$ NPs becomes larger and form $Cu_2O$ nanocubes as those evidences in the dashed yellow square in Figure 9(b). Passing to the image of Figure 9(c), at $t_w$=5 min the size of the $Cu_2O$ nanocubes continue to increase. This observation marks a clear difference with the evolution of the Cu2O nanocubes formed on the inner chromatic rings whose sizes keep pretty stable passing from sample B to E (see Figs. 8(b)–8(e)). Moreover, in this last case the Cu film appears almost eroded and larger regions of uncovered substrate surface emerge in the outer annulus F$_{OA}$. The Cu$_2$O NPs/nanocubes in the three regions of D$_{OA}$, E$_{OA}$ and F$_{OA}$ are rather scattered spatially and easily identified, which makes possible to get a statistically reliable measure of their size distribution. Figure 9(d)–9(f) report the size histograms of the Cu$_2$O NPs/nanocubes, while the upper-right insets in Figure 9(d)–9(f) show the particles selected for the size distribution measurement. In particular, the selected points are 218 in the region D$_{OA}$, 144 in E$_{OA}$ and 114 in F$_{OA}$. The Cu$_2$O NPs in D$_{OA}$ are characterized by an average size of ≈20 nm, with the largest fraction (≈60%) of NPs falling in the 15–25 nm size bin. The average size of the Cu$_2$O nanocubes in E$_{OA}$ and F$_{OA}$ are ≈30 nm and ≈60 nm, respectively, which clearly evidences the progressive growth of the Cu$_2$O nanocubes dimension with the increase of the immersion time.

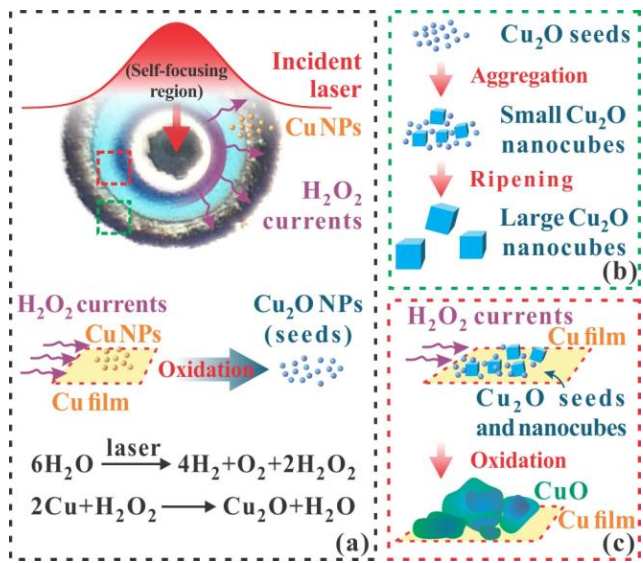

**Figure 10.** Schematic illustration of the processes leading to Cu$_2$O nanocube and irregular CuO particle formations. (a) Generation of the initial Cu$_2$O NPs (Cu$_2$O seeds) described by the reactions

at the bottom of (a). The developing processes in (b) and (c) correspond to the regions marked by the dashed squares in red and green, respectively.

Figure 10 schematically depicts a proposed scenario to rationalize our experimental findings and address chemical reactions and aggregation processes leading to the formation of $Cu_2O$ nanocube and irregular CuO particle during water immersion. Panel (a) of Figure 10 illustrates the initial stage of the process generating $Cu_2O$ NPs, which successively act as seeds to form $Cu_2O$ nanocubes, and the involved chemical reactions. It is worth observing that at the focus the laser beam intensity is enhanced by the self-focusing effect of water, which undergoing ionization and dissociation at intensities between $10^{13}$ and $10^{16}$ W/cm$^2$ leads to the formation of hydrogen peroxide ($H_2O_2$).[41] Moreover, Cu NPs are produced during the laser irradiation on the Cu film in liquid water.[42,43] Current flows generated in the liquid will push around both Cu NPs and $H_2O_2$ moving them off the center of the laser beam focus, which accounts for the ≈5 $\mu$m intervals between the laser-induced crater and the chromatic annuli observed in the OM micrographs of Figure 5. In the meantime, the Cu NPs and the Cu on the surface of the film will be oxidized into $Cu_2O$ by reaction with $H_2O_2$ according to the reactions depicted in the bottom of Figure 10(a).[44,45] resulting in the formation of the ultrafine $Cu_2O$ seeds. In Figure 10, panels (b) and (c) illustrates the processes occurring in the two regions marked by dashed green and red squares, respectively, in Figure 10(a). Let's consider first the outer black annulus (dashed green square in Figure 10(a) and Figure 10(b)). As this region is at larger distance from the laser beam focus, it is likely characterized by a lower fraction of oxidizing agents than the inner chromatic annulus and a lower density of Cu NPs. As a result, the $Cu_2O$ NPs seeds (≈20 nm) aggregate and form the small $Cu_2O$ nanocubes (≈30 nm) observed in the SEM image in Figure 9(b) for $t_w$=4 min. Then, at longer immersion time, the small $Cu_2O$ nanocubes ripen to generate large $Cu_2O$ nanocubes ($\approx$60 nm), observed in the SEM images in Figure 9(c).[25] As for the inner cyan annulus marked by the dashed red square in Figure 10(a), this is located in the area characterized by a more oxidizing environment and with a larger density of Cu NPs. Hence, it is possible that the small $Cu_2O$ nanocubes cannot ripen to large ones but more likely form irregular CuO particles by reacting with the oxidizing agents ($H_2O_2$, $O_2$, etc.). As the irradiation time increases, the surface of the Cu thin film is progressively more covered by CuO particles or $Cu_2O$/CuO mixture particles, and less material is left that can still undergo oxidization processes. As a consequence, this will result in the rather negligible variation of the sizes of the oxidized copper nanostructures with the immersion time observed for the inner annuli, as observed in the evolution of the morphology depicted by the SEM images of Figure 8. In conclusion, the water provides a better circumstance than air for the generation of copper oxides particles with lager size and $Cu_2O$ nanocrystal with particular shape. We address that it is the tiny amount of $H_2O_2$ generated by the self-focusing effect of water that plays an important role in the oxidation processes, and finally leads to the generation of CuO particles, $Cu_2O$/CuO mixture particles and $Cu_2O$ nanocubes.

## 4. Conclusion

We investigated high repetition rate fs laser irradiation of copper thin films, both in air and liquid water. Our approach demonstrate novel striking effects of laser induced oxidization of Cu thin films in such a regime. We observe the formation of different chromatic annuli which is likely related to the high repetition rate periodic heating of the sample and to the non-uniform energy distribution induced by heat diffusion along the thin film plane. The irradiated samples are characterized by optical and SEM microscopy and $\mu$Raman spectroscopy. Laser processing in air leads to the formation of chromatic copper oxides annuli (CuO and $Cu_2O$) around the laser beam focus location that extend up to ≈100 $\mu$m, at 1 min irradiation time. Interestingly, a striking dependence of the characteristic features of the laser-induced oxidation process on the copper thin film thickness $H$ (namely, ≈50 nm, ≈100 nm and ≈200 nm) is addressed, which shows that $Cu_2O$ can effectively formed aver a large area for $H$≈200 nm, while distinct annular regions of CuO and to $Cu_2O$ are formed for smaller values of $H$.

As for the processing in water, the seldom analyzed effect of immersion time in the liquid, after an irradiation with a fixed number of laser shots, is investigated. We observe the clear formation of chromatic annuli also in this case, whose color and composition changes with the immersion time. In particular, $Cu_2O$ NPs (size of ≈20 nm) and $Cu_2O$ nanocubes (nanocube edges of ≈30, ≈60 nm) can be effectively produced with an immersion time of few minutes. An interpretation scenario is proposed to explain the laser-induced oxidation processes leading to our experimental findings. Our experimental findings pave a novel, interesting way for CuO NPs and $Cu_2O$ NPs/nanocubes fabrications with potentiality for gas sensing, photovoltaic materials and catalyst photo-activation applications, e.g., which takes advantages by the peculiar characteristics of processing with high repetition rate femtosecond laser sources.

**AUTHOR INFORMATION**

**Corresponding Author**

*E-mail: huminglie@tju.edu.cn.

**Present Addresses**

†Ultrafast Laser Laboratory, Key Laboratory of Opto-electronic Information Technical Science of Ministry of Education, College of Precision Instruments and Opto-electronics Engineering, Tianjin University, Tianjin 300072, China

‡Dipartimento di Fisica, Università di Napoli Federico II, Complesso Universitario di Monte S. Angelo, Via Cintia, I-80126 Napoli, Italy

**ACKNOWLEDGMENT**

S.H. thanks the China National Scholarship Fund. M.H. acknowledges National Natural Science Foundation of China (NSFC) (Grant Nos. 61322502, 61535009, and 11274239), and the Program for Changjiang Scholars and Innovative Research Team in University (Grant No. IRT13033).